\documentclass[]{spie}  
\usepackage[]{graphicx}

\title{An Atmospheric Dispersion Corrector Design with Milliarcsecond-Level Precision from 1 to 4 microns for High Dispersion Coronagraphy} 

\author{Jason J. Wang\supit{a}, J. Kent Wallace\supit{b}, Nemanja Jovanovic\supit{a}, Olivier Guyon\supit{c,d,e}, Mitsuko Roberts\supit{a}, and Dimitri Mawet\supit{a,b}
\skiplinehalf
\supit{a}Department of Astronomy, California Institute of Technology, Pasadena, CA 91125, USA; \\
\supit{b}Jet Propulsion Laboratory, California Institute of Technology, 4800 Oak Grove Dr.,Pasadena, CA 91109, USA \\
\supit{c}Subaru Telescope, National Astronomical Observatory of Japan, 650 North Aohoku Place, Hilo, HI 96720, USA \\
\supit{d}Steward Observatory, University of Arizona, Tucson, AZ 85721, USA \\
\supit{e}Astrobiology Center of NINS, 2-21-1 Osawa, Mitaka, Tokyo 181-8588, Japan
}

\authorinfo{Further author information: (Send correspondence to J.J.W.)\\J.J.W.: E-mail: jwang4@caltech.edu}

\begin{document} 
  \maketitle 

\begin{abstract}
Differential atmospheric refraction (DAR) limits the amount of light that can be coupled into a single mode fiber and provides additional complications for any fiber tracking system. We present an atmospheric dispersion corrector (ADC) design based off of two counter-rotating prisms to fit the needs of exoplanet spectroscopy for the Keck Planet Imager and Characterizer (KPIC) from 1.1 to 4.2 microns. Due to strong telluric effects, we find that the default Zemax prescription for DAR between 2 and 4.2 microns to be inaccurate up to 15 mas when comparing against DAR models computed from first principles. Using first-principle models, we developed our own custom ADC optimization solution and achieve less than 4 mas residual dispersion in any individual science band (J, K, L) down to 60 degree zenith angles, while the whole time maintaining less than 3 mas of residual dispersion in the tracking band (H) and less than 2 mas of residual dispersion between the tracking and science bands.
\end{abstract}



\section{INTRODUCTION}
\label{sec:intro}  
The Keck Planet Imager and Characterizer (KPIC)\cite{Mawet2016,Mawet2018} feeds light from the Keck II adaptive optics (AO) system into the NIRSPEC infrared spectrograph using single mode fibers. This has the advantage of spatially isolating the light of the planet from the bright glare of the star before spectrally isolating it with a spectrograph, in a technique known as high dispersion coronagraphy (HDC)\cite{Wang2017,Mawet2017}. HDC techniques enable detailed characterization of exoplanets with measurements of planetary radial velocity, planetary spin, and atmospheric properties\cite{Snellen2014, Brogi2019}, some of which cannot be measured through any other means. 

Due to the high-dispersion nature, HDC depends on collecting as many photons as possible on the detector in order to have enough planet photons in each spectral channel to study the planet's atmosphere. The primary challenge along this front for KPIC is coupling in as much planetary photons as possible into the single mode fibers that feed the spectrograph. The KPIC fiber injection unit (FIU) is designed to do exactly this using the light that has been corrected by the AO system at Keck. The current Phase I version of the KPIC FIU is fairly basic, consisting of a tip-tilt mirror that steers light in x-y to guide it into the fiber and using the Keck AO deformable mirror to perform static non-common path aberration correction \cite{jovanovic2019}. 

KPIC Phase II consists of a series of upgrades to the FIU that will drastically improve its capabilities to maximize fiber coupling \cite{Pezzato2019, calvin2020-EDS, jovanovic2020-PVC}. In this work, we will focus on the atmospheric dispersion corrector (ADC). The ADC is designed to correct for differential atmospheric refraction (DAR), the dispersion of light as a function of wavelength due to Earth's atmosphere acting as a prism. DAR causes the apparent position of any astrophysical source to change with wavelength, which results in there being no placement of the fiber in the focal plane that can optimize coupling into a single mode fiber at all wavelengths of interest. Thus, ADCs serve an important role in counteracting the dispersion of the atmosphere before light is focused into the fibers. Due to the small angular size of single mode fibers (essentially the diffraction limit of a point source), the KPIC ADC must correct DAR to milliarcsecond-level precision in order to optimize coupling into the single mode fibers. 

\section{Model of Differential Atmospheric Refraction}
Designing an ADC to meet milliarcsecond-level requirements requires not only tight optical requirements, but also a precise knowledge of differential atmospheric refraction (DAR). The amount of differential atmospheric refraction depends the index of refraction of air right above the primary mirror of the telescope under the plane parallel atmosphere approximation. The amplitude can be written as:
\begin{equation}
    \delta z = \left( \frac{n^2 - 1}{2n^2} - \frac{n_0^2 - 1}{2 n_0^2} \right) \tan(z),
\end{equation}
where $\delta z$ is the amplitude (in radians) of DAR at a zenith angle $z$ for a wavelength with index of refraction $n$ compared to a reference wavelength with index $n_0$. The key is computing $n$, which for milliarcsecond precision in the near-infrared, requires eight significant figures of accuracy. 

\begin{figure}
    \begin{center}
    \includegraphics[width=0.90\textwidth]{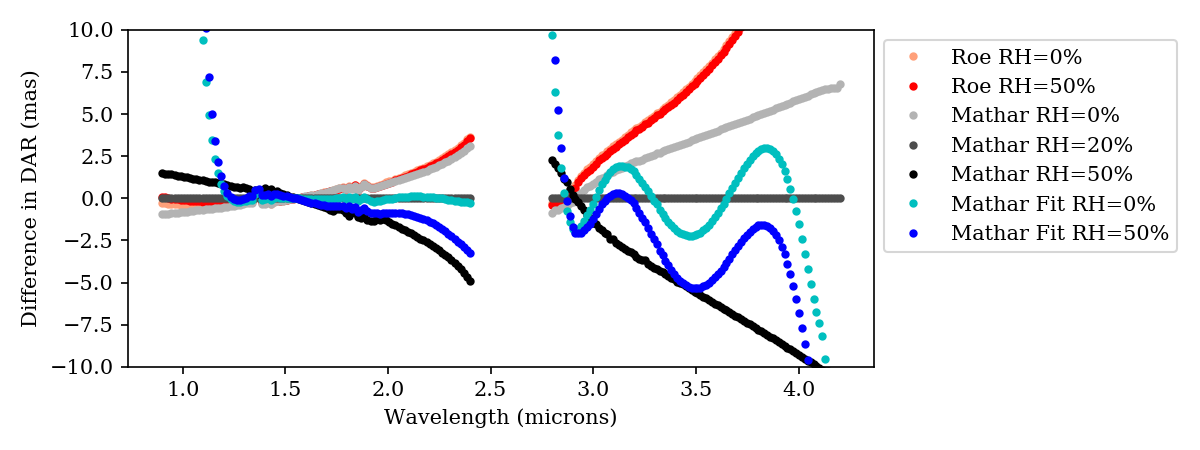}
    \end{center}
   \caption[example] 
   { \label{fig:atm_models} Difference between various models of the index of refraction of humid air. The reference model is the Mathar model with 20\% humidity. The Roe models, which are approximately what is in Zemax, and the Mathar Fit models differ significantly in the $K$ and $L$-bands where the effect of humidity is the strongest.}
\end{figure} 

We considered three models for $n$ from the literature. First, we considered a simple model (which we call``Roe") that is valid from the visible to the infrared from Refs \citenum{Schubert2000} and \citenum{Roe2002}. We believe this model is nearly the same model used in the Zemax optical modeling software. Second, we considered the polynomial approximation presented in Ref \citenum{Mathar2007} (which we call ``Mathar Fit") that is valid in certain wavelength ranges (e.g., 1.3-2.5 $\mu$m and 2.8-4.2 $\mu$m). Lastly, we computed our own grid of $n$ from first-principles using the publicly available code that was used in Ref \citenum{Mathar2007} (which we call ``Mathar"), and used linear interpolation to approximate $n$ between grid points. This last model has been benchmarked at $N$-band and shown to be consistent with empirical measurements \cite{Skemer2009}. We plot a comparison of the DAR predicted by these three classes of models in Figure \ref{fig:atm_models}. For all the models, we used a temperature of 276.15~K, pressure of 61400~Pa, and a zenith angle of $60^\circ$. 

We see that in the $K$ and $L$-bands, the disagreements are quite significant. In the case of 50\% relative humidity, the values can differ by over 10~mas. The poses a problem for designing a milliarcsecond-level precision ADC, as its performance will depend on what the actual index of air is at a given pressure, temperature, and relative humidity. Clearly more observations are needed to identify the right model for DAR. In this work, we will adopt the Mathar first-principles model since it is not an approximation and has at least been validated at longer wavelengths for astronomical applications. We further verified this model is consistent with another first-principles model currently in development down to the 1~mas level. 

\begin{figure}
    \begin{center}
    \includegraphics[width=0.90\textwidth]{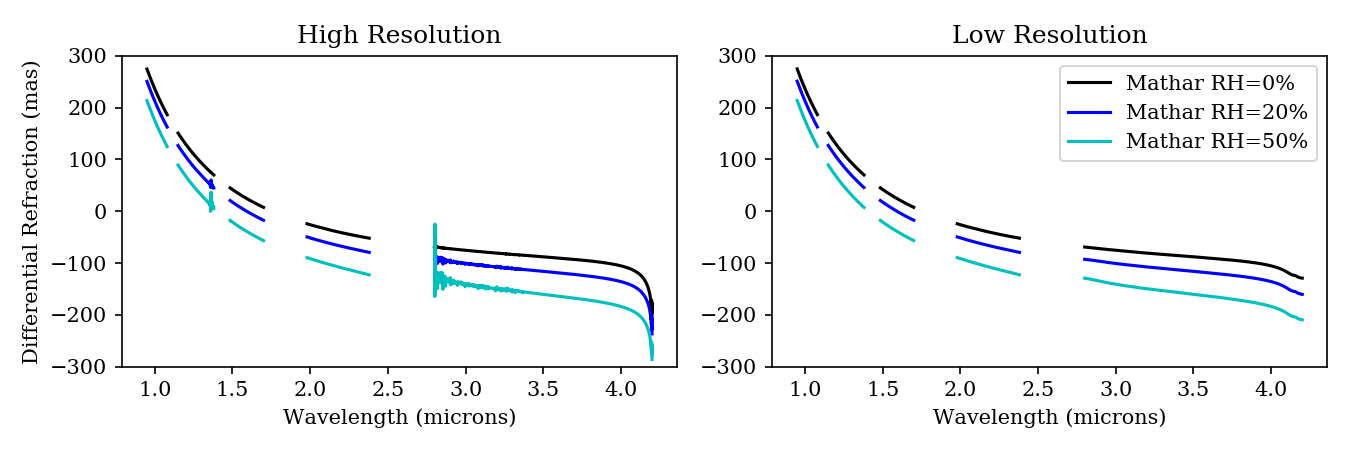}
    \end{center}
   \caption[example] 
   { \label{fig:dar_raw} The effect of DAR as a function of wavelength for different levels of humidity. The plot on the left is the full effect of DAR, whereas the plot on the right is smoothed down to R < 100 to remove spikes in DAR that we have ignored.  }
\end{figure}

For the bands of interest in this paper, we used the Mathar model and computed the peak-to-valley (PTV) dispersion in each astronomical band and the median offset in dispersion in that band from $H$-band in Table \ref{tab:dar_vanilla}. We plot the full profile due to DAR in Figure \ref{fig:dar_raw}. In areas with strong telluric absorption, high spectral frequency spikes in DAR occurs. These spikes are not able to be corrected by our ADC design, and occur at wavelengths where most astrophysical light is absorbed. Thus, these regions are not the focus for us scientifically. We have smoothed out and ignored these spikes when designing our ADC (right panel of Figure \ref{fig:dar_raw}. 
   
\begin{table}
\caption{The effect of uncorrected differential atmospheric refraction in median conditions} 
\label{tab:dar_vanilla}
\begin{center}       
\begin{tabular}{|c|c|c|c|} 
\hline
Band & Wavelengths ($\mu$m) & PTV Dispersion in Band (mas) & Offset from $H$-band (mas) \\
\hline
J & 1.150-1.380 & 81.1 & 81.0 \\
\hline
H & 1.485-1.810 & 38.9 & 0.0 \\
\hline
K & 1.980-2.380 & 29.8 & 65.9 \\
\hline
L & 2.950-3.950 & 33.5 & 114.4 \\
\hline
\end{tabular}
\end{center}
\end{table} 

\section{Optical Design} 

\begin{figure}
    \begin{center}
    \includegraphics[width=0.50\textwidth]{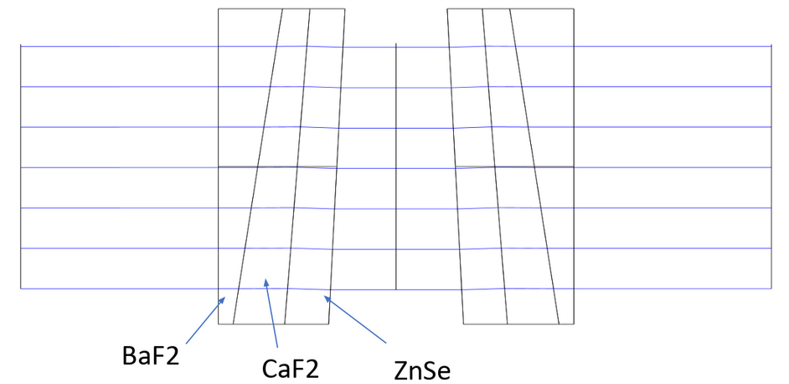}
    \end{center}
   \caption[example] 
   { \label{fig:adc_Zemax} Optical diagram of the ADC design. The three materials that comprise the prisms are labeled. }
\end{figure}

Our ADC design consists of two counter-rotating prisms, with each prism consisting of three wedges made of BaF2, CaF2, and ZnSe. This is a similar design as that presented in Ref.~\citenum{Kopan2013} for visible wavelength high-angular resolution imaging. The prism is designed to operate on a collimated beam. The orientation in Figure \ref{fig:adc_Zemax} induces the maximum possible dispersion. The second prism can be rotated $180^\circ$ about the optical axis to cause no net dispersion when light passed through both prisms. In the design phase, the free parameters are the angles of each of the three wedges that make up each prism. During operation, the one free parameter is the relation rotation between the two prisms (termed the ``clocking angle") which controls the magnitude of the net dispersion caused by the prism pair. 

Due to the choice of the Mathar first principles model for the index of refraction of humid air, we were not able to optimize the design of the ADC in Zemax. Instead, we wrote a custom Python code to model this ADC design, and validated it produces the same numerical results if given the same input dispersion as Zemax. 

\subsection{Design Requirements} 
\label{sec:rec}
The dispersion compensation requirements of the ADC are listed in Table \ref{tab:req}. The three requirements are the peak-to-valley (PTV) dispersion in the science band of interest, the median dispersion between the science band and $H$-band for tracking, and the PTV dispersion in the $H$-band. For all three cases, the PTV dispersion in $H$-band is set to ensure the PSF on the tracking camera is not elongated. The $K$- and $L$-band requirements for PTV dispersion in the science band and offset from $H$-band are set by the science for direct spectroscopy of exoplanets, and in particular maximizing fiber coupling of the on-axis planet light into the fiber. In both of these cases the requirement is 0.1 $\lambda/D$ for both parameters. The $J$-band requirements are set by sensitivity requirements for the Vortex Fiber Nulling (VFN)\cite{Echeverri2019, Echeverri2020}, with tolerances of 0.01 $\lambda/D$ for both the PTV dispersion in the $J$-band and the median offset between $J$ and $H$ bands. 

\begin{table}
\caption{Design Requirements for the KPIC ADC} 
\label{tab:req}
\begin{center}       
\begin{tabular}{|c|c|c|c|c|} 
\hline
Band & Max Zenith Angle & PTV Dispersion in Band & Offset from $H$ & PTV Dispersion in $H$-band \\
 & (deg) & (mas) & (mas) & (mas) \\
\hline
J & 30 & 0.3 & 0.3 & 3 \\
\hline
K & 60 & 4 & 4 & 3 \\
\hline
L & 60 & 8 & 8 & 3 \\
\hline
\end{tabular}
\end{center}
\end{table}

\subsection{Optimization}\label{sec:opt}
The physical design of our ADC has three free parameters: the angles of each of three wedges in the prisms. In addition, the clocking angle of the ADC can be different in each observing mode, since our ADC does not need to meet the requirements in $J$, $H$, and $K$ simultaneously. For example, a band that will require a large dispersion compensation may require one clocking of the ADC, but a band with significantly less dispersion may want a different clocking angle to achieve the right amount of dispersion compensation. 

We fixed the index of refraction of air based on the median conditions on Maunakea: temperature of 276.15 K, pressure of 61400 Pa, and a relative humidity of 20\%. We explored optimization on more extreme cases where the DAR is the largest (e.g., 50\% relative humidity) but found that it worsened ADC performance in conditions at low humidity levels. We used the max zenith angle listed in Table \ref{tab:req} as the zenith angle to optimize each band with. Smaller zenith angles are simply handled by changing the clocking angle of the ADC. 

We set up a cost function with 12 terms. For each of the three science bands, we add to the cost function the PTV dispersion in the science band, the median dispersion between the science and tracking band, and the PTV dispersion in the tracking band. This results in 9 terms. The last term is to minimize the angular deflection of the of the $H$-band beam leaving the ADC in each of the three configurations. This is to simplify optical alignment so that we can assume both the incoming and outgoing beams are nearly perpendicular to the front and back surfaces of the ADC. 

We explored the cost function space using the \texttt{emcee} sampler\cite{ForemanMackey2013}. We note that this analysis is not Bayesian, but we used it to simply understand the covariances between the parameters. We found that the three prism angles are highly covariant with each other. Regardless, we took the solution that minimized our cost function as the optimal parameters for our ADC. The optimal parameters for the wedge angles are listed in \ref{tab:params}. 

\begin{table}
\caption{Optimized Parameters for the ADC} 
\label{tab:params}
\begin{center}       
\begin{tabular}{|c|c|c|} 
\hline
BaF2 Wedge Angle &  CaF2 Wedge Angle & ZnSe Wedge Angle \\
 (deg) & (deg) & (deg) \\
\hline
7.0516 & 3.8050 & 1.1465 \\
\hline
\end{tabular}
\end{center}
\end{table} 

\section{Performance}

\subsection{Comparison to Design Requirements}
We optimized the performance of our ADC against the median Maunakea atmospheric conditions specified in Section \ref{sec:opt}. We quantified performance of our optimized ADC in Table \ref{tab:perf} against the parameters defined in Section \ref{sec:rec}. Our ADC design performs better than all three requirements in all three bands for all three requirements. We achieved less than 4~mas of residual dispersion in any science band, while maintaining less than 3~mas of dispersion in the tracking band and less than 2 mas of median dispersion between the science and tracking bands. With this ADC design, KPIC will be able to maximize coupling of planet light down to $60^\circ$ zenith angles. 

However, this assumes a single temperature, pressure, and humidity in the atmosphere and assumes perfect manufacturing and control. In the remaining sections, we will assess the sensitivity of our performance to changes in the atmosphere and imperfections in the manufacturing and control. 

\begin{table}
\caption{ADC Performance Relative to Design Requirements (listed in parentheses)} 
\label{tab:perf}
\begin{center}       
\begin{tabular}{|c|c|c|c|c|c|} 
\hline
Band & Zenith Angle & Clocking Angle & PTV Dispersion in Band & Offset from $H$ & PTV Dispersion in $H$ \\
 & (deg) & (deg) & (mas) & (mas) & (mas) \\
\hline
J & 30 & 74.1892 & 0.22 (0.3) & 0.14 (0.3) & 0.21 (3) \\
\hline
K & 60 & 34.1697 & 0.23 (4) & 0.28 (4) & 0.16 (3) \\
\hline
L & 60 & 39.1265 & 3.20 (8) & 1.26 (8) & 2.52 (3) \\
\hline
\end{tabular}
\end{center}
\end{table} 

\begin{figure}
    \begin{center}
    \includegraphics[width=0.47\textwidth]{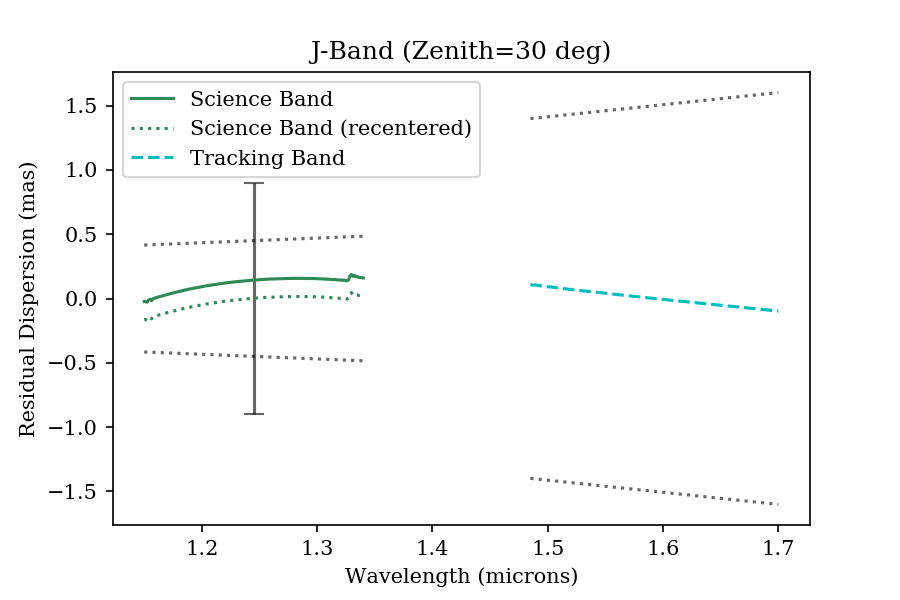}
    \includegraphics[width=0.47\textwidth]{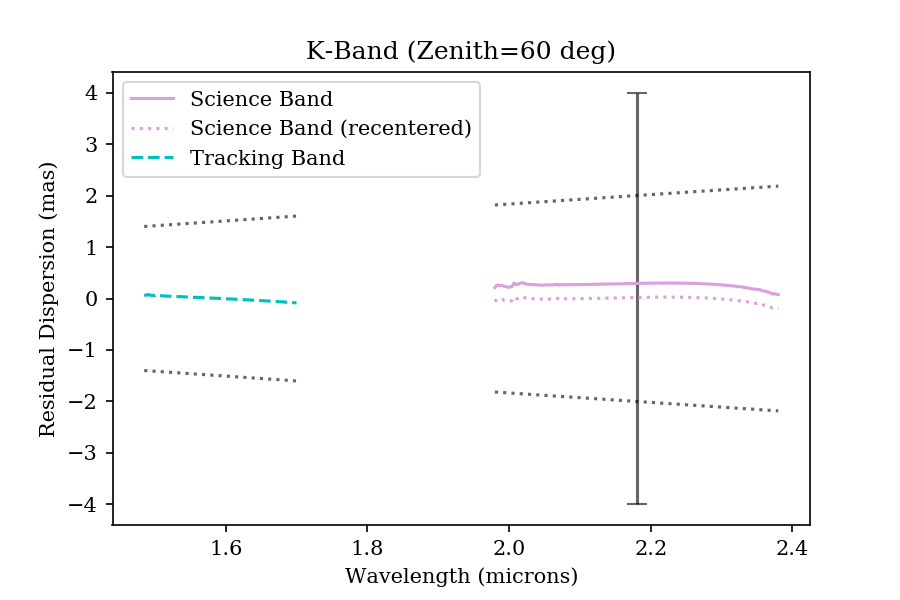}
    \includegraphics[width=0.47\textwidth]{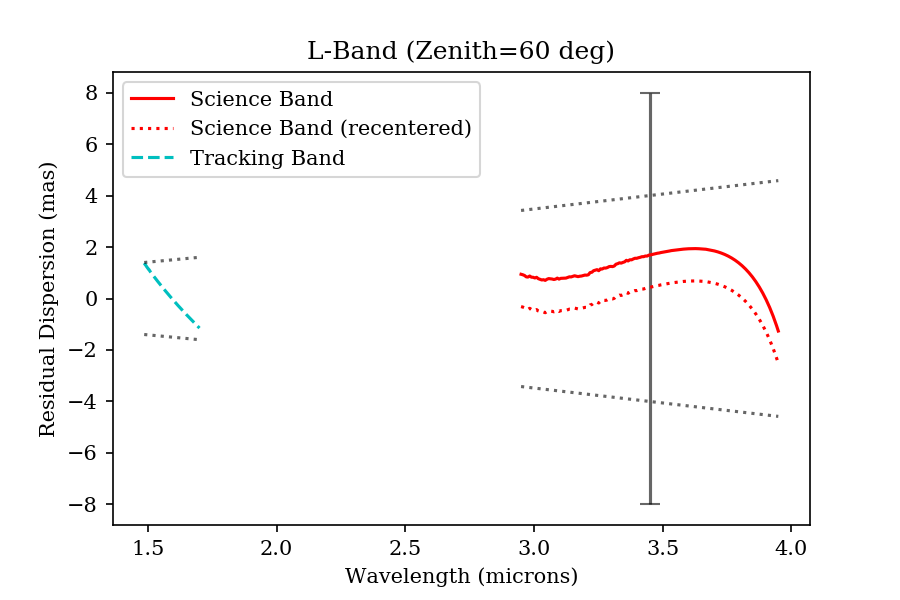}
    \end{center}
   \caption[example] 
   { \label{fig:perf} ADC performance in the three different observing modes and wavelengths. The dashed line in the science band assumes the median offset is compensated by offsetting the fiber position from the reference $H$-band position.}
\end{figure} 
   
\subsection{Sensitivity to Atmospheric Conditions}
We used historical Maunkea weather data to assess the variance in temperature, pressure, and relative humidity on the summit. We approximated the real distributions with the following approximations for numerical convenience. For pressure, we used a Gaussian distribution centered at 61400 Pa with a standard deviation of 1850 Pa. For temperature, we used a Gaussian centered at $3^\circ$ C with a sigma of $2.4^\circ$ C. For relative humidity, we used an uniform distribution between 0\% and 50\%. 

\begin{table}
\caption{Sensitivity of ADC performance to changes in atmospheric conditions. For each quantity, the 95\% value is listed. Only 5\% of cases are worse than this value. Requirements listed in parentheses. } 
\label{tab:atm_perf}
\begin{center}       
\begin{tabular}{|c|c|c|c|c|} 
\hline
Band & Zenith Angle & 95\% PTV Dispersion in Band & 95\% Offset from $H$ & 95\% PTV Dispersion in $H$ \\
 & (deg) & (mas) & (mas) & (mas) \\
\hline
J & 30 & 0.210 (0.3) & 0.279 (0.3) & 0.45 (3) \\
\hline
K & 60 & 1.76 (4) & 2.16 (4) & 1.46 (3) \\
\hline
L & 60 & 6.90 (8) & 11.21 (8) & 4.00 (3) \\
\hline
\end{tabular}
\end{center}
\end{table} 

To assess our ADC design against changes in atmospheric parameters, we kept the same ADC wedge angles in Table \ref{tab:params} that were optimized for the median conditions, and only changed the clocking angle to adapt to difference atmospheric conditions. We simulated random atmospheric parameters drawn from the distributions specified above, optimized the ADC clocking angle to best correct for DAR, and quantified how well it met requirements in each of the 3 bands. 

In Table \ref{tab:atm_perf}, we list the 95-\%tile performance for each of the requirements, which means that 95\% of the time, the performance is at least as good as this. In $J$ and $K$ bands, we meet the requirements even when varying atmospheric values in realistic ways. In $L$-band, the offset from $H$-band and the dispersion in the tracking band marginally exceed requirements, indicating that $L$-band performance is most sensitive to atmospheric conditions. 

\begin{figure}
    \begin{center}
    \includegraphics[width=0.97\textwidth]{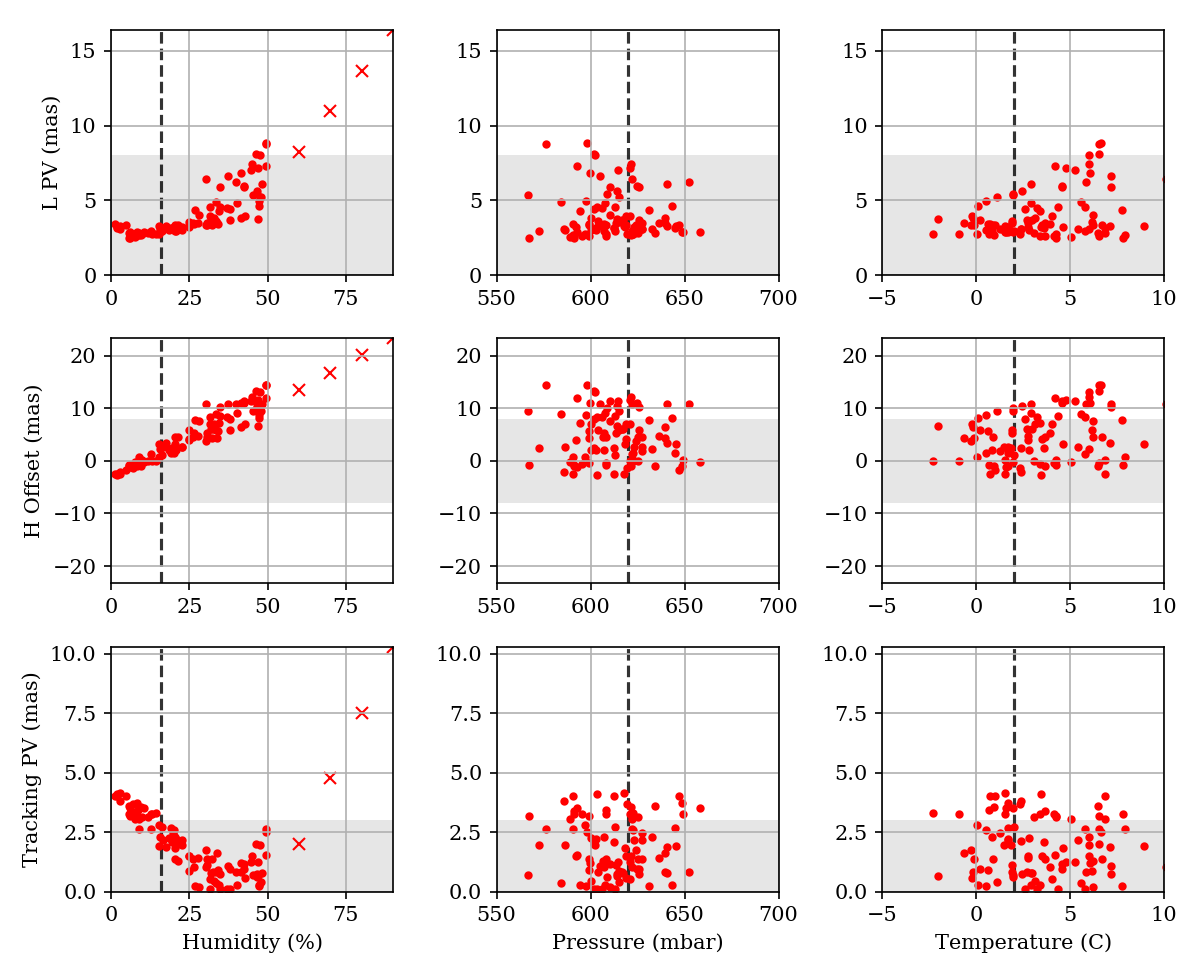}
    \end{center}
   \caption[example] 
   { \label{fig:lband_perf} Performance of the ADC in $L$-band against 100 realizations of atmospheric conditions. Median atmospheric parameters are plotted with the vertical dashed line. The gray shaded region indicates where requirements are satisfied. Red x markers show extreme humidity cases. Strong correlations of performance with relative humidity are seen.}
\end{figure}

In Figure \ref{fig:lband_perf}, we further investigated $L$-band sensitivity to atmospheric conditions by plotting the three ADC performance metrics against each draw of temperature, pressure, and humidity that we considered. We see strong correlateions of ADC performance with relative humidity. In particular, PTV dispersion in the science band and offset from $H$-band tracking gets worse as relative humidity increases, as one might expect. However, PTV dispersion in the tracking band gets worse at low humidity. As PTV dispersion in the science band cannot be corrected in any other way, we are satisfied that we meet performance in nearly all cases for this metric. The offset from $H$-band can be compensated for by offsetting a source from the nominal position determined by the $H$-band tracking camera by the appropriate amount. Thus, fiber coupling can still be maximized even with an offset above requirement between the science and tracking bands. The marginal elongation of the PSF in the tracking band at low humidity should not significantly impact performance, based on our experience in KPIC Phase I, and can be compensated for by fitting for an elongated PSF in our tracking operations.

\subsection{Sensitivity to Wedge Angle}
It is impossible to construct wedges with angles exactly as we simulated, so we assess the performance of our ADC with errors in the wedge angles. We assumed we will be able to measure the manufactured wedge angles in the lab, and thus adapt our control script to account for the actual wedge angles. We assumed all 6 wedges (3 wedges per prism) have their own manufacturing error. For each wedge, we randomly drew a manufacturing error from a Gaussian distribution centered at 0 with a standard deviation of $0.01^\circ$. The standard deviation was chosen based on evaluations from manufacturers who said such a tolerance can be met. We adopt the median atmospheric conditions that we used in optimization, and only changed the clocking angle of the ADC to compensate for the changes in wedge angles. 

\begin{table}
\caption{Sensitivity of ADC Performance to Errors in Wedge Angle. The format is the same as Table \ref{tab:atm_perf}. } 
\label{tab:wedge_perf}
\begin{center}       
\begin{tabular}{|c|c|c|c|c|} 
\hline
Band & Zenith Angle & 95\% PTV Dispersion in Band & 95\% Offset from $H$ & 95\% PTV Dispersion in $H$ \\
 & (deg) & (mas) & (mas) & (mas) \\
\hline
J & 30 & 0.12 (0.3) & 0.27 (0.3) & 0.28 (3) \\
\hline
K & 60 & 0.24 (4) & 0.47 (4) & 0.16 (3) \\
\hline
L & 60 & 3.28 (8) & 2.71 (8) & 2.61 (3) \\
\hline
\end{tabular}
\end{center}
\end{table} 

In Table \ref{tab:wedge_perf}, we list the 95-\%tile performance for the three science bands. We meet requirements on all counts, even for 2$\sigma$ deviations. Thus, we can more than tolerate $0.01^\circ$ manufacturing errors in the wedge angles, assuming the actual angles are characterized well in the lab. 

\subsection{Sensitivity to Clocking Angle}
The control of the ADC requires rotating the prisms to the appropriate orientation in order to produce the exact amount of dispersion to cancel out DAR. The ADC prisms are mounted on rotating stages with encoders that can move the stages in discrete steps. If the steps are too big, we would not be able to rotate the stages precisely enough for optimal dispersion compensation. We investigated this by simulating the ADC using the optimal wedge angles and median atmospheric parameters, but adding quantization errors to the requested clocking angle. 

\begin{figure}
    \begin{center}
    \includegraphics[width=0.97\textwidth]{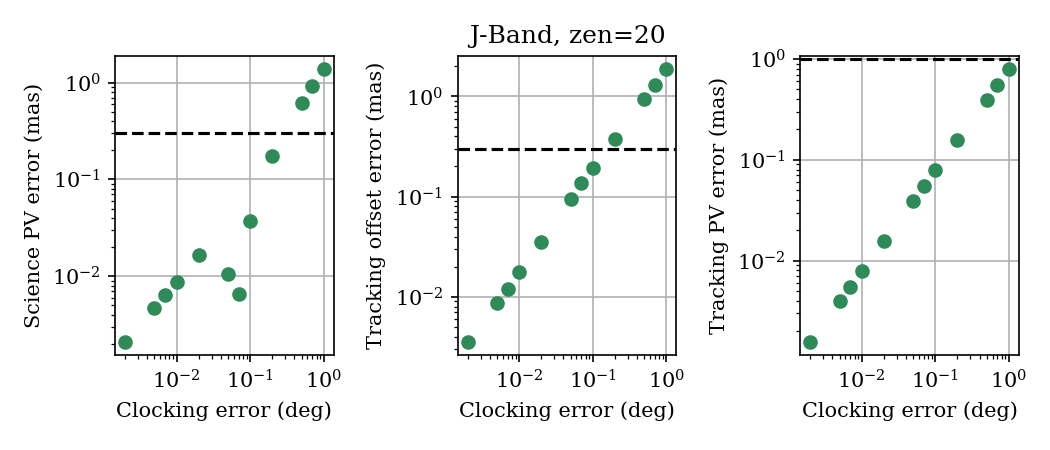}
    \end{center}
   \caption[example] 
   { \label{fig:jband_clock} Performance of the ADC in $J$-band as a function of clocking angle quantization error. 1/3 of the requirement value is marked by the dashed horizontal line. A clocking angle of $0.2^\circ$ at most is tolerable to keep clocking errors below 1/3 of the maximum allowed dispersion. }
\end{figure}

We found that the $J$-band VFN mode is the most sensitive to clocking angle requirements. We plot the performance of the ADC in $J$-band as a function of clocking angle quantization error for a $20^\circ$ zenith angle in Figure \ref{fig:jband_clock}. We found that the clocking stage needs to have a precision of $0.2^\circ$ or better to keep quantization errors well below the requirement (1/3 of the requirement value). In $K$ and $L$-bands, even a precision of $1^\circ$ for the clocking angle rotation stage is sufficient.

\section{Conclusion}
We presented the design, optimization, and sensitivity analysis for the ADC for KPIC. In simulation, our ADC is able to achieve milliarcsecond-level dispersion control in $J$, $K$, and $L$ bands simultaneous with $H$-band. The broadband ability of our ADC will allow for multiple different science cases for KPIC \cite{jovanovic2020-PVC, Echeverri2020}. 

The biggest remaining area of concern for our ADC is the uncertainty in models for DAR, and the fact the current models have not been validated empirically at these wavelengths at this level of precision. In particular, $L$-band is the most sensitive to changing atmospheric conditions, and but changes in DAR due to changes in humidity have not been measured empirically. As DAR becomes even more important for future thirty-meter-class telescopes with even higher angular resolution, it will be important to empirically measure the index of refraction of humid air in the infrared at these levels of precision.

\acknowledgments     
This work was supported by the Heising-Simons Foundation through grants \#2019-1312 and \#2015-129. J. Wang is supported by the Heising-Simons Foundation 51 Pegasi b postdoctoral fellowship. Part of this work was carried out at the Jet Propulsion Laboratory, California Institute of Technology, under contract with the National Aeronautics and Space Administration (NASA). W. M. Keck Observatory is operated as a scientific partnership among the California Institute of Technology, the University of California, and the National Aeronautics and Space Administration (NASA). The Observatory was made possible by the generous financial support of the W. M. Keck Foundation. The authors wish to recognize and acknowledge the very significant cultural role and reverence that the summit of Maunakea has always had within the indigenous Hawaiian community. We are most fortunate to have the opportunity to conduct observations from this mountain.


\bibliography{report}   
\bibliographystyle{spiebib}   

\end{document}